\documentclass[12pt]{iopart}

\usepackage{graphicx}
\usepackage{nicefrac}
\usepackage{hyperref}
\usepackage{braket}
\usepackage{mathdots}
\usepackage{enumitem}

\usepackage{iopams}

\begin{document}

\title[On the quantum behavior of correlated portfolios]{On the quantum behavior and clustering properties of correlated financial portfolios}
\author{C. R. da Cunha and  R. da Silva}
\address{Instituto de F\'isica, Universidade Federal do Rio Grande do Sul, 91501-970 Porto Alegre, Rio Grande do Sul, Brazil.}
\ead{creq@if.ufrgs.br}

\begin{abstract}
	We investigate 17 digital currencies making an analogy with quantum systems and develop the concept of ``eigenportfolios''. We show that the density of states of the correlation matrix of these assets shows a behavior between that of the Wishart ensemble and one whose elements are Cauchy distributed. A metric for the participation ratio based on the superposition of Gaussian functions is proposed and we show that small eigenvalues correspond to localized states. Nonetheless, some level of localization is also present for bigger eigenvalues probably caused by the fat tails of the distribution of returns of these assets. We also show through a clustering study that the digital currencies tend to stagger together. We conclude the paper performing ultrametric ordering of the assets and show as an stylized fact that the market tend to value significantly distinctive assets.
\end{abstract}


\noindent
\submitto{\EJP}

\maketitle

\section{Introduction}

The idea of finding parallels between quantum mechanics and financial markets has attracted the attention of scientists and economists for
many years. Early attempts to find those parallels include a quantum model for binomial markets\cite{QuantBin} and the capture of the famous Black-Scholes\cite{BS} model within a quantum physical setting\cite{BSQuantum,QuantumBS}. Others have tried to use a path integral formulation\cite{PathInt,PathInt2,PathInt3} or even quantum game theory\cite{quantumGame} to simulate the pricing of options. More recently, quantum algorithms have been proposed for the pricing of financial derivatives\cite{QComp} and even the explicit concept of quantum economics has been proposed \cite{QMoney}. Similar connections between physical and economic systems have been defining the field of Econophysics \cite{Grass}. 

Quantum theory is undoubtedly very successful in treating electronic systems with strong correlations.
Here investigate some of the tools used in quantum theory to study  the correlated structure of financial markets.
Our findings led us to believe that portfolios of 
currencies resemble quantum mechanical systems. Also 
localization properties of their eigenmodes suggest
a strong and interesting connection between stock markets and quantum
mechanics. 

We test these concepts on digital currencies.
These are financial assets deprived of an issuing agent
constructed on a peer-to-peer managed distributed ledger platform called
blockchain. Consensus about the trading of these assets is decentralized,
thus they resemble physical autonomous agents that operate out of
equilibrium but often reach some steady state regimes where financial returns are characterized by some
power law \cite{Mantegna}. For instance,
Bitcoin \cite{Bitcoin}, the most popular digital currency has attracted the
attention of the scientific community despite much criticisms. For example, it has been shown that Bitcoin shares a set of stylized facts
with standard financial instruments and also mimics many natural phenomena\cite{US}.

Since the development of Bitcoin in 2008, many other digital coins appeared
in the market as a result of successive forks of its main code. This has
happened mainly as a tendency for differentiation in order to make the asset
more suitable for specific purposes in a clear case of Darwinian
adaptation \cite{Darwin}. There are over 4000 different digital coins
available today but only a few have a significant market cap that can attract the attention
of investors.

We divide the remaining of the paper as follows. In section \ref{Section:Our_data} we present
the data used in this paper as well as some fundamental background 
connecting random matrices, crypto markets, and correlations. We
also describe the notation to be used along the paper. In the following
section (section \ref{Section:Results}) we present a brief theoretical formulation and our main results separated by context  in a few subsections. 
Furthermore, we show that these coins have a natural tendency of clustering. We investigate this phenomena studying the minimum spanning tree formed by 
connecting these assets according to their spectral properties. Finally, we calculate  the ultrametric ordering of the assets and show that the market tends to value more cryptocurrencies that are significantly distinctive.
Finally our summaries and conclusions are briefly explained in
section \ref{Section:Conclusions}.

\section{Description of Data and Definitions}
\label{Section:Our_data}

The historical series for $M=20$\ digital assets were directly downloaded
from the HitBTC application programming interface (https://api.hitbtc.com)
for $N=10^{5}$\ minutes ending on 2019-08-12. Three currencies with
insufficient historical data were dropped and a portfolio was formed with $%
M=17$\ assets: BTC, ETH, EOS, LTC, BCH, TRX, ETC, XMR, NEO, XEM, IOTA, XLM,
DASH, ZEC, ADA, DOGE, and XTZ. The logarithmic returns for the closing price
of these digital coins were calculated as:

\begin{eqnarray}
	r_{m,n}		 &= \log (p_{m,n+1})-\log (p_{m,n})\nonumber \\
						& \approx  \frac{p_{m,n+1}}{p_{m,n}}-1,
\end{eqnarray}
where $p_{m,n}$ is the $n^{\rm{th}}$ closing price for the $m^{\rm{th}}$
digital asset, with $n=1...N$ spanning the length of the time series ($10^5$ minutes), and $m=1...M$ spanning the number of assets (17). We then computed the
normalized returns that form a matrix $\mathbf{L}$ with elements:

\begin{equation}
	L_{m,n}=\frac{r_{m,n}-\langle r_{m}\rangle }{\sqrt{\left\langle	r_{m}^{2}\right\rangle -\langle r_{m}\rangle ^{2}}},
	\label{Eq:L}	
\end{equation}
where, for any observable $O$ in this work, we assign: 
\begin{equation}
	\langle O_{m}\rangle =\frac{1}{N}\sum_{n=1}^{N}O_{m,n}.
\end{equation}

The crypto market is composed of only a few major crypto currencies
that dominate the marketshare. Moreover, most of the alternate
cryptocurrencies are constructed on already existing blockchains, which
restricts the size of the matrices that can be analyzed. This creates a major problem for many spectral analyses that are used in this work. Specifically in our
case, there are only $M=17$\ assets in the portfolio, which may not produce
adequate statistics. To circumvent this problem, we constructed an ensemble of
1000 correlation matrices:

\begin{equation}
	\boldsymbol{\Lambda }=\frac{1}{N}\mathbf{L}'\mathbf{L}'^{\rm{T}},
	\label{Eq:correlation}
\end{equation}
with elements related to Eq. \ref{Eq:L}:

\begin{equation}
\Lambda _{ij}=\frac{\langle r_{i}r_{j}\rangle -\langle r_{i}\rangle \langle
	r_{j}\rangle }{\sqrt{\left\langle \left( \Delta r_{i}\right)
		^{2}\right\rangle }\sqrt{\left\langle \left( \Delta r_{j}\right)
		^{2}\right\rangle }},
\end{equation}
where $\mathbf{L}'$ is a randomly resampled matrix from the original matrix $\mathbf{L}$ obtained with replacement taking different windows of $N=100$ continuous closing prices. Also, $\left\langle\left(\Delta r_i\right)^2\right\rangle=\left\langle\left(r_i-\langle r_i\rangle\right)^2\right\rangle=\langle r_i^2\rangle-\langle r_i\rangle^2$.
The dimension of the $\mathbf{L}'$ matrices is 17 $\times$ 100. It is important to note that if the elements of $\mathbf{L}$ were also independent and identically distributed (\textit{i.i.d}) Gaussian random variables, then $\boldsymbol{\Lambda}$ would compose the so-called Wishart ensemble\cite{Wishart}. This is a key concept since the density of eigenvalues for this particular case is analytically known. Therefore, we can use this result to compare corresponding matrices $\boldsymbol{\Lambda}$ with those obtained with experimental data.

\section{Formulation and results: Quantum Analogies and network properties in the context of cryptocurrencies}
\label{Section:Results}
Our main results are presented in this section. For every subsection, we perform a brief theoretical formulation with a subsequent presentation of the numerical results and data analysis.

\subsection{Quantum Mechanics and the concept of Eigenportfolios}

An unobserved quantum system is supposed to have a finite number $N$ of elements in a particular ensemble. A state of this physical system in a Hilbert space is then represented by a ket $\ket{\varphi}$ in Dirac notation. This state can be decomposed in a linear combination kets $\ket{m}$ where $m\in[1;N_{ind}]$:

\begin{equation}
	\ket{\varphi}=\sum_{m=1}^{N_{ind}}p_m\ket{m},
	\label{eq:Upla}
\end{equation}
where $p_m$ is the relative population of the state $\ket{m}$. This coefficients are obviously non-negative and add up to the unity, i.e. $\sum_{m=1}^Np_m=1$. It is important to point out that $N_{ind}$ is the number of elements in the ensemble, not the dimension of the space generated by the eigenvectors of a certain physical observable. There is, actually, a great arbitrariness on the choice of kets $\ket{m}$. For instance, they can freely chosen to be orthogonal or not, or to be the eigenstates of some particular operator or even the eigenstates of two operators simultaneously. Considering the ensemble, if there is only one n-upla $(p_1,\dots,p_N)$ that satisfies Eq. \ref{eq:Upla}, then the ensemble is considered to be pure. Otherwise, if there is more than one n-upla $(p_1,\dots,p_N)$ that satisfies Eq. \ref{eq:Upla}, then the ensemble is considered to be mixed.

Thus, in order to perform a measurement of some observable $\hat{O}$, one must perform averages over the ensemble of such observable. This is given by:

\begin{equation}
	\left\langle\hat{O}\right\rangle = \sum_{m=1}^{N_{ind}}p_m\braket{m|\hat{O}|m}.
\end{equation}

Using the completeness relation $\sum_l\ket{l}\bra{l}=\mathbf{1}$, where $\ket{l}$ are eigenstates of $\hat{O}$, i.e. $\hat{O}\ket{l}=l\ket{l}$, we obtain:

\begin{eqnarray}
		\left\langle\hat{O}\right\rangle &= \sum_{m=1}^{N_{ind}}p_m\braket{m|\hat{O}\mathbf{1}|m}\nonumber\\
			&= \sum_{m=1}^{N_{ind}}\sum_lp_m\braket{m|\hat{O}|l}\braket{l|m}\nonumber\\
			&= \sum_{m=1}^{N_{ind}}\sum_llp_m\braket{m|l}\braket{l|m}\nonumber\\
			&= \sum_{m=1}^{N_{ind}}\sum_llp_m|\braket{m|l}|^2.
\end{eqnarray}

This procedure can be performed twice introducing a new basis:

\begin{eqnarray}
		\left\langle\hat{O}\right\rangle &= \sum_{m=1}^{N_{ind}}p_m\braket{m|\mathbf{1}\hat{O}\mathbf{1}|m}\nonumber\\
			&= \sum_{l,k}\sum_{m-1}^{N_ind}p_m\braket{k|m}\braket{m|l}\braket{l|\hat{O}|k}\nonumber\\
			&= \sum_{l,k}\rho_{kl}O_{lk},
\end{eqnarray}
where $O_{lk}=\braket{l|\hat{O}|k}$ and $\rho_{kl}=\sum_{m=1}^{N_{ind}}p_m\braket{k|m}\braket{m|l}$ are  elements of the hermitian operator $\hat{\rho}$ known as \emph{density operator}:

\begin{equation}
	\hat{\rho}=\hat{\rho}^\dagger=\sum_{m=1}^{N_{ind}}p_m\ket{m}\bra{m}.
\end{equation}

If one is interested to study the time evolution of the statistical ensemble, the time derivative of the density operator must be taken:

\begin{equation}
	\frac{\partial\hat{\rho}}{\partial t} = \sum_{m=1}^{N_{ind}}p_m\left(\frac{\partial\ket{m}}{\partial t}\bra{m}+\ket{m}\frac{\partial\bra{m}}{\partial t}\right).
\end{equation}
Substituting $i\hbar\frac{\partial}{\partial t}\rightarrow H$, one obtains the evolution of the density matrix, which is given be a quantum version of the Liouville equation:

\begin{equation}
	i\hbar\frac{\partial\hat{\rho}}{\partial t} = [H,\hat{\rho}].
\end{equation}

In steady state, the density matrix commutes with the Hamiltonian. Therefore, they share a common eigenbasis and we can use it to formulate a quantum description of the crypto market. For this purpose, we apply the Rayleigh-Ritz variational principle to find the ground state of the correlation operator $\boldsymbol{\hat{\Lambda}}$. In order to do so, we expand the correlation matrix in the basis $\ket{n}$ of its eigenstates:

\begin{equation}
	\boldsymbol{\hat{\Lambda}} = \sum_{n=0}^{M-1}\lambda_n\ket{n}\bra{n}.
\end{equation}

Therefore, the elements of the correlation matrix can be written as:

\begin{eqnarray}
		\braket{W|\boldsymbol{\hat{\Lambda}}|W} &= \sum_n\lambda_n\braket{W|n}\braket{n|W}\nonumber\\
			&\geq\lambda\sum_{n=0}^M\braket{W|n}\braket{n|W}\nonumber\\
			&\geq\lambda\braket{W|\left(\sum_{n=0}^M\ket{n}\bra{n}\right)|W}=\lambda\braket{W|W},
\end{eqnarray}
where $\lambda=\min\{\lambda_n\}$ corresponds to the ground state of the system. In order to find ground state, the following difference has to be minimized:

\begin{equation}
	F = \bra{W}\boldsymbol{\hat{\Lambda}}\ket{W}-\lambda\braket{W|W}.
\end{equation}
This is the same expression one would obtain minimizing the volatility $\mathbf{W}^\textrm{T}\boldsymbol{\Lambda}\mathbf{W}$ for a normalized portfolio $\mathbf{W}^\textrm{T}\mathbf{W}=1$.
The extremum of this function is given by:

\begin{equation}
	\boldsymbol{\Lambda}\mathbf{W}=\lambda \mathbf{W}.
	\label{eq:eigenp}
\end{equation}
In order to explore the quantum behavior of the portfolios we will use $\boldsymbol{\hat{\Lambda}}=\boldsymbol{\Lambda}$ as given by Eq. \ref{Eq:correlation}.

The ``\emph{eigenportfolio}'' $\mathbf{W}$ (or \emph{principal component}) is on the eigenbasis $\ket{n}$ of the
correlation matrix, whereas the normalized eigenvalues (or \emph{characteristic roots}), i.e. $\lambda_j'=\lambda
_{j}/\sum_{k=0}^{M-1}\lambda _{k}$, give the explained volatility of the $j^{\rm{th}}$\ portfolio.  
This is exactly what one would obtain from a principal component analysis (PCA)\cite{PCA} of the assets.
On the other hand, had the Lagrangian function $F$ included a term that imposes a specific
value to the expected return, one would obtain the so-called Markowitz
Lagrangian used in the modern portfolio theory (MPT)\cite{Markowitz}.

It is interesting to compare the obtained explained volatility
with two limiting cases in order to study the experimental data. One case is obtained from the Wishart ensemble considering the matrix $\mathbf{L}$ previously defined in section \ref{Section:Our_data} and used to construct $\boldsymbol{\Lambda}$. The elements of $\mathbf{L}$ are distributed according to a normal distribution. The Wishart ensemble has been used for the spectral analysis of the density matrix in general quantum systems\cite{Xiao}.
The other limiting case was obtained from matrices $\mathbf{L}$ whose elements are \textit{i.i.d.} random variables drawn from the Cauchy distribution:

\begin{equation}
	f(x;\gamma )=\frac{1}{\pi \gamma }\left[ \frac{\gamma ^{2}}{x^{2}+\gamma ^{2}%
	}\right] ,
\end{equation}
where $\gamma $ is a scale parameter that indicates the half-width at
half-maximum of the distribution. This distribution was particularly chosen
because it is the only known stable pathological distribution on a support $(-\infty,+\infty)$ (both positive and negative returns have to be accounted for) that has fat tails and can be expressed analytically.

\begin{figure}[h]
	\centering
	\includegraphics{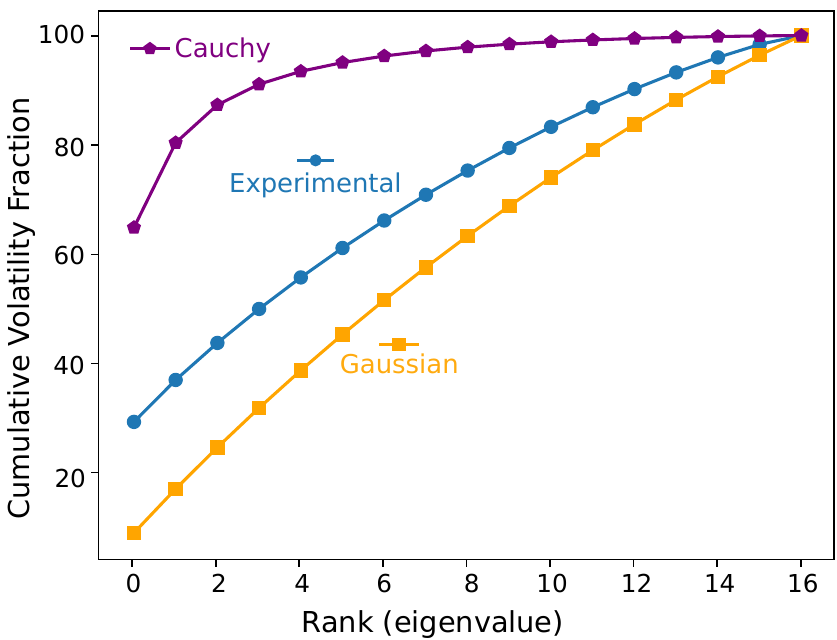}
	\caption{Expected cumulative volatility fraction explained as a function of the rank number
	of eigenvalues for the experimental data, and control series with Gaussian
	and Cauchy distributions with $\gamma=0.1$ (triangle marks) and $\gamma=0.45$ (pentagon marks).}
	\label{fig:PCA}
\end{figure}

The cumulative volatility fraction shown in Fig. \ref{fig:PCA} is defined as:

\begin{equation}
	\phi(j)=\sum_{m=0}^j\lambda_m'=\frac{\sum_{k=0}^j\lambda_k}{\sum_{k=0}^{M-1}\lambda_k},
\end{equation}
where $j$ is the rank of the eigenvalue $\lambda_j$.The figure indicates that the largest eigenvalue explains
approximately 30 \% of the variance. Moreover, the cumulative volatility is
located between those of Gaussian and Cauchy ensembles. Thus, we discard the hypothesis that the
distribution of returns come from a Wiener process. However, the results also indicate that the hypothesis that the
distribution has infinite moments may be discarded. The result indicates that, more likely, the distribution of returns
has long tails, as previously obtained  for Bitcoin\cite{US}.

\subsection{Random Matrices and Density of States}

Given this initial formulation we now turn our attention to some properties of cryptocurrencies in the context of random matrices, a concept widely used in quantum chaos.
Eugene Wigner has originally proposed in 1955 that the energy levels of heavy nuclei
could be described by the eigenvalues of random matrices. A Wigner matrix
is then defined as a square matrix that is symmetric (but in more general cases
it can also be Hermitian or symplectic), where their elements $%
\{h_{ij}|\forall i\leq j\}$ and $\{h_{ii}\}$ are identically distributed random variables
according to a symmetric distribution with finite first and second moments. Following this condition we can obtain the
joint distribution of its eigenvalues, and integrating all values except
one leads to a universal semicircle law for the the density eigenvalues \cite{Wigner}.

Nonetheless, the semi-circle law does not apply to our matrix of returns since it is non-square. Rather, one can use the Mar\v{c}enko-Pastur
law \cite{MP} to analyze and compare the density of eigenvalues of the $M\times M$
correlation matrices for $N$ returns.
If the elements of $L$ are mutually
uncorrelated \textit{i.i.d.} standard normal random variables, the Mar\v{c}enko-Pastur law states that the density of eigenvalues of $\boldsymbol{\Lambda}$ is given by:

\begin{equation}
	\rho (\lambda )=\frac{N}{2\pi M}\frac{\sqrt{(\lambda -\lambda _{\min
		})(\lambda_{\max} -\lambda)}}{\lambda },  
	\label{Eq:Marchenko}
\end{equation}%
where 
\begin{equation}
	\lambda _{\max ,\min }=\left( 1\pm \sqrt{\frac{N}{M}}\right) ^{2}.
	\label{eq:Lambdas}
\end{equation}

Thus, it is possible to verify whether the density of eigenvalues of $\boldsymbol{\Lambda}$ follows the law described by Eq. \ref{Eq:Marchenko}. Unlike directly computed by other
authors (see for example \cite{Plerou}), we calculate the
density of eigenvalues of $\boldsymbol{\Lambda}$ using the elegant method of Green's
function for two reasons: 

\begin{enumerate}[label=\roman*),font=\itshape]
	\item Our correlation matrix has low dimensionality. Thus, we would run the risk of using insufficient bins for obtaining a meaningful histogram for the density of eigenvalues;
	\item This approach makes the formalism used in quantum mechanics explicit to treat  economic problems found in this new field of Econophysics.
\end{enumerate}

\subsubsection{Green's functions}

The retarded Green's function associated with the correlation matrix $%
\boldsymbol{\Lambda }$ is given by the resolvent:

\begin{equation}
	\mathbf{G}_\eta^{\rm{R}}(\lambda)=\left[(\lambda+i\eta)\mathbf{I}-\boldsymbol{\Lambda}\right]^{-1},
\end{equation}
where $\eta\rightarrow0^+$. Considering $\ket{m}$ as an eigenvector of $\boldsymbol{\Lambda}$ with eigenvalue $\lambda_m$ we have:

\begin{eqnarray}
		\mathbf{G}_\eta^R(\lambda)\ket{m}&=(\lambda+i\eta)^{-1}\left[\mathbf{I}-\frac{\boldsymbol{\Lambda}}{\lambda+i\eta}\right]^{-1}\ket{m}\nonumber\\
			&= (\lambda+i\eta)^{-1}\sum_{j=0}^\infty\frac{1}{(\lambda+i\eta)^j}\boldsymbol{\Lambda^j}\ket{m}\nonumber\\
			&= (\lambda+i\eta)^{-1}\left(\sum_{j=0}^\infty\frac{1}{(\lambda+i\eta)^j}\lambda_m^j\right)\ket{m}\nonumber\\
			&= (\lambda+i\eta)^{-1}\left(1-\frac{\lambda_m}{\lambda+i\eta}\right)^{-1}\ket{m}\nonumber\\
			&= \frac{1}{(\lambda-\lambda_m)+i\eta}\ket{m},
\end{eqnarray}
where we used the geometric series representation of the operator $\left[(\lambda+i\eta)\mathbf{I}-\boldsymbol{\Lambda}\right]^{-1}$ as $(\lambda+i\eta)^{-1}\sum_{j=0}^\infty\boldsymbol{\Lambda}/(\lambda+i\eta)^j$.

Thus, using the closeness relation, one obtains the Green's function written in the eigenbasis of the correlation matrix:

\begin{eqnarray}
		\mathbf{G}_\eta^R(\lambda) &= \mathbf{G}_\eta^R(\lambda)\mathbf{1}\nonumber\\
			&= \sum_{m=0}^{M-1}\left[(\lambda+i\eta)\mathbf{I}-\boldsymbol{\Lambda}\right]^{-1}\ket{m}\bra{m}\nonumber\\
			&= \sum_{m=0}^{M-1}\ket{m}\frac{1}{(\lambda-\lambda_m)+i\eta}\bra{m}\nonumber\\
			&= \sum_{m=0}^{M-1}\ket{m}\mathbf{C}_m,
\end{eqnarray}
where:

\begin{equation}
	\mathbf{C}_{m}=\frac{\bra{m}}{(\lambda -\lambda_m)+i\eta }.
\end{equation}
Therefore, the retarded Green's function can be written as:

\begin{equation}
	\mathbf{G}_{\eta }^{\rm{R}}(\lambda )=\sum_{m=0}^{M-1}\frac{\ket{m}\bra{m}}{(\lambda-\lambda_m) +i\eta}.
\end{equation}

Taking the limit $\eta\rightarrow 0^+$, one obtains:

\begin{equation}
	\lim_{\eta \rightarrow 0^+}\rm{Im}\left\{\bra{n}\mathbf{G}_{\eta }^{\rm{R}}\ket{n}\right\} =-\pi \delta (\lambda -\lambda_{n}).
\end{equation}%
Thus we can calculate the density of states according to:

\begin{eqnarray}
		\rho (\lambda ) &= \frac{1}{M}\sum_{n=0}^{M-1}\delta (\lambda -\lambda _{n})\nonumber\\
					  		  &=  -\frac{1}{\pi M}\lim_{\eta \rightarrow 0}\textrm{Tr}\left\{ \rm{Im}
							\left\{ \mathbf{G}_{\eta }^{\rm{R}}(\lambda )\right\} \right\}.
	\label{Eq:Density}
\end{eqnarray}

Again, we average it over an ensemble of $n_{samp.}=$1000 matrices to get
a mean density of assets $\bar{\rho}(\lambda )=\left\langle \rho
(\lambda )\right\rangle $.

\begin{figure}[h]
	\centering
	\includegraphics{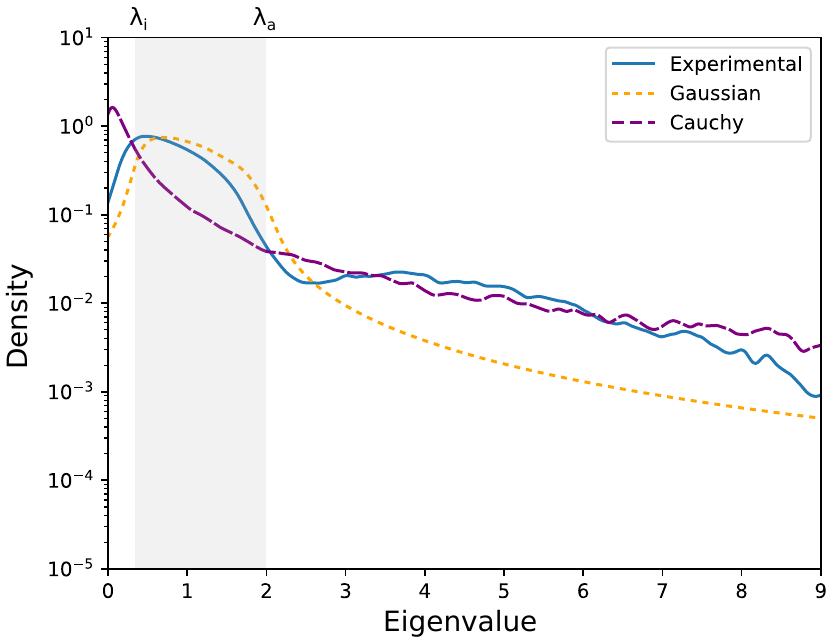}
	\caption{Measured density of eigenvalues. The shaded area corresponds to the
	region between the expected minimum and maximum eigenvalues from the Mar\v{c}%
	eko-Pastur distribution (Eq. \protect\ref{eq:Lambdas}. The density of
	control series with Gaussian and Cauchy distributions are also shown.}
	\label{fig:MP}
\end{figure}

The density obtained for the series under study in Fig. \ref{fig:MP} resembles that of Gaussian ensembles. However, it occupies a region of eigenvalues slightly different than that expected from the Mar\v{c}enko-Pastur distribution in Eq. \ref
{eq:Lambdas}. The expected theoretical distribution should fit a region between $\lambda _{\min}=0.34$ and $\lambda_{\max}=1.99$. However, our data occupies a region between $\lambda _{\min}=0.03$ and $\lambda _{\max}=11.73$. 
A least square fitting with a Cauchy distribution for the elements of the correlation matrix was computed to obtain the density of eigenvalues closest to the experimental data. Our calculation produced a scale parameter $\gamma=0.032$. Also, the observed shift as well as the increase in the density values for the tail obtained experimentally seem to be related to the presence of non-Gaussian returns in the portfolio. Similar analysis of the density using random matrix theory on daily stock prices of the Tokyo Stock Exchange produced similar deviations from the Mar\v{c}enko-Pastur distribution for smaller eigenvalues\cite{Utsugi}. However, the deviations were attributed to different levels of randomness.

\begin{figure}[h]
	\centering
	\includegraphics{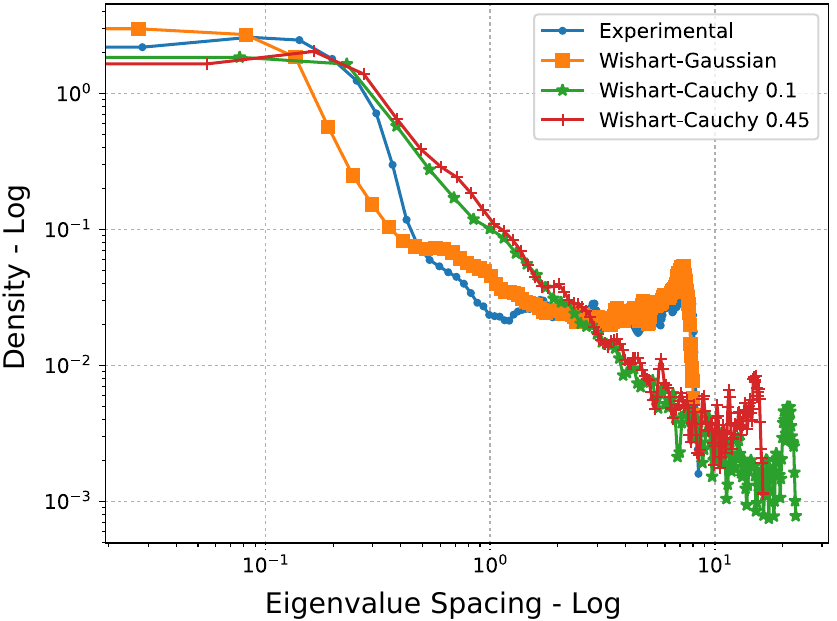}
	\caption{Eigenvalue spacing distributions for the experimental data as well as the benchmark distributions.}
	\label{fig:Surm}
\end{figure}

Although the elements of a Gaussian ensemble are \textit{i.i.d}, their eigenvalues are not. Rather, any two eigenvalues of a Gaussian matrix are unlikely to be close to each other in value. This is known as \emph{level repulsion}.
We investigated this behavior by calculating the cumulative density of eigenvalues $F(\lambda)=\int_{-\infty}^\lambda\rho(y)dy$ and drawing $10^4$ uniform random values $u\in[0;1]$ such that $\lambda=F'^{-1}(u)$ where $F'^{-1}(u)$ is a linear interpolation for $F(\lambda)^{-1}$. The eigenvalues $\lambda$ were then unfolded. For this we calculated $\eta(\lambda)=\sum_n\theta(\lambda-\lambda_n)$ and averaged $\eta(\lambda)$ over an ensemble of 100 random sets of $\lambda$.  The spacing of these estimated levels were then calculated (only for nearest neighbors) and their histograms are shown in Fig. \ref{fig:Surm} for the experimental data, the Wishart-Gaussian ensemble, and for Wishart-Cauchy ensembles with scale parameters of $\gamma=0.032$ and $\gamma=0.048$. The position of the experimental data in Fig. \ref{fig:Surm} between the benchmark distributions reinforce our hypothesis that some elements of the portfolio have non-Gaussian distributions.

Having studied the density
of eigenvalues and benchmarked it, we will now discuss the quantum localization phenomena applied to
cryptocurrencies. 

\subsection{Measures of Localization}

Although the portfolio is normalized, its fourth moment is empirically related to the
localization of the eigenfunctions and is commonly known as the \emph{%
inverse participation ratio }\cite{partip1}:

\begin{equation}
	\Delta(\lambda _{m})=\left(\sum_{k=1}^{M}|\braket{e_k|m}|^{4}\right)^{-1},
	\label{eq:particip}
\end{equation}%
where $\ket{m}$ ($m\in[0;M-1]$) are the eigenvectors of $\boldsymbol{\Lambda}$ and $e_{k}$ ($k\in[0;M-1]$) forms a complete canonical orthonormal basis in a Hilbert space with dimension $M$.

It is interesting to consider some cases to illustrate this concept. For instance,
an eigenvector with equal components $\braket{e_k|m}=M^{-\nicefrac{1}{2}}$
has an $\Delta=M$ (the dimension of $\boldsymbol{\Lambda}$), whereas an eigenvector with only one component
different than zero $\braket{e_k|m}=\delta_{mm^{\prime}}$ has $\Delta=1$.
Therefore, the inverse participation ratio gives the number of modes
in a specific eigenvector. Here we explore the concept of normalized participation ratio,
where $\Delta=1$ indicates an \emph{extended state} and $\Delta=M^{-1}$ indicates a \emph{localized state}.

\begin{figure}[h]
	\centering
	\includegraphics{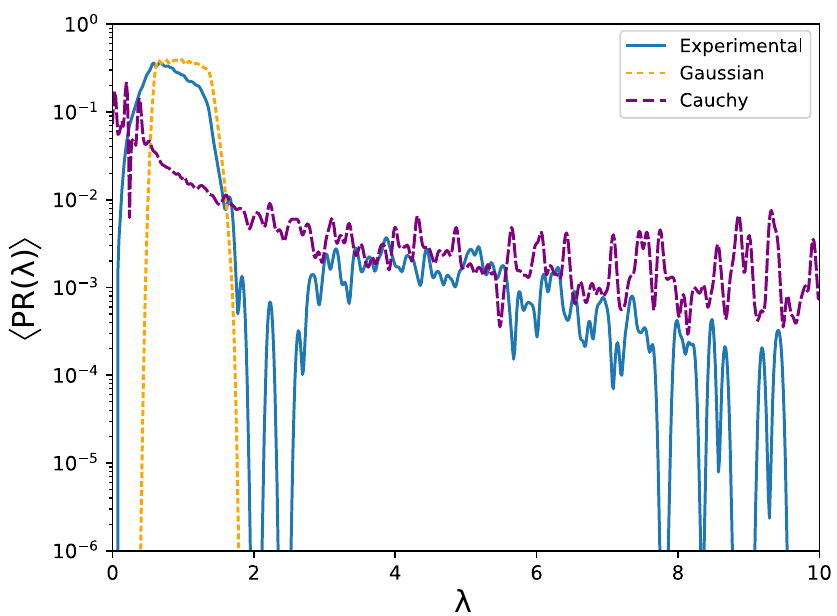}
	\caption{Expected participation ratio $\langle PR(\protect\lambda)\rangle$
	obtained for: the time series under study (blue), a Gaussian control
	(orange), and a Cauchy control with scale parameter $\protect\gamma=0.0481$
	(purple).}
	\label{fig:Particip}
\end{figure}

In order to obtain $\Delta$ as a continuous function and average it over the whole time series we adopted the
following fitting procedure. First, we construct a function $\varphi(\lambda)$ with Gaussian
functions centered on each eigenvalue:

\begin{equation}
	\varphi (\lambda )=\sum_{m=0}^{M-1}A_{m}\exp \left( -\frac{(\lambda -\lambda_{m})^{2}}{2\sigma ^{2}}\right),
\end{equation}
where $A_{m}$ are constants to be determined. Here, from an \emph{ad hoc} point of
view, we can consider this function to assume the value of the participation
ratio for each eigenvalue:

\begin{equation}
	\varphi (\lambda _{k})=\Delta(\lambda _{k})=\sum_{m=1}^{M}A_{m}\exp \left( -%
	\frac{(\lambda _{k}-\lambda _{m})^{2}}{2\sigma ^{2}}\right).
\end{equation}
This produces a set of equations that can be put in a matrix format:

\begin{equation}
	\mathbf{EA}=\boldsymbol{\Delta},
\end{equation}
where $\mathbf{E}$ has elements given by:

\begin{equation}
	E_{ab} = \exp\left(-\frac{(\lambda_a-\lambda_b)^2}{2\sigma^2}\right),
\end{equation}
and $\boldsymbol{\Delta}$ has elements given by $\Delta(\lambda_a)$ calculated from Eq. \ref{eq:particip}.

This can be solved to find the coefficients $A_{n}$ and construct the
function $\varphi (\lambda)$, which will be taken as $\Delta(\lambda)$. This function is then calculated and averaged
for different regions of the time series to produce an expected
participation ratio $\langle\Delta(\lambda)\rangle $.

We calculated the expected participation ratio for the experimental data and for the Wishart-Gaussian ensemble as shown in Fig. \ref{fig:Particip}. We also applied least squares fitting to find a Wishart-Cauchy distribution whose participation ratio that best fitted that of experimental data. In our case, the scale parameter found was $\gamma=0.048$.
The expected participation ratio for the experimental data indicates the presence of extended states
for small eigenvalues as found for the Wishart-Gaussian ensemble \cite{partip2}. The shift of localization for very small eigenvalues as well as the mild localization found for larger eigenvalues, though, can again be attributed to the presence of elements with non-Gaussian distributions in the portfolio.

\subsection{Correlation Structure}

\begin{figure}[h]
	\centering
	\includegraphics{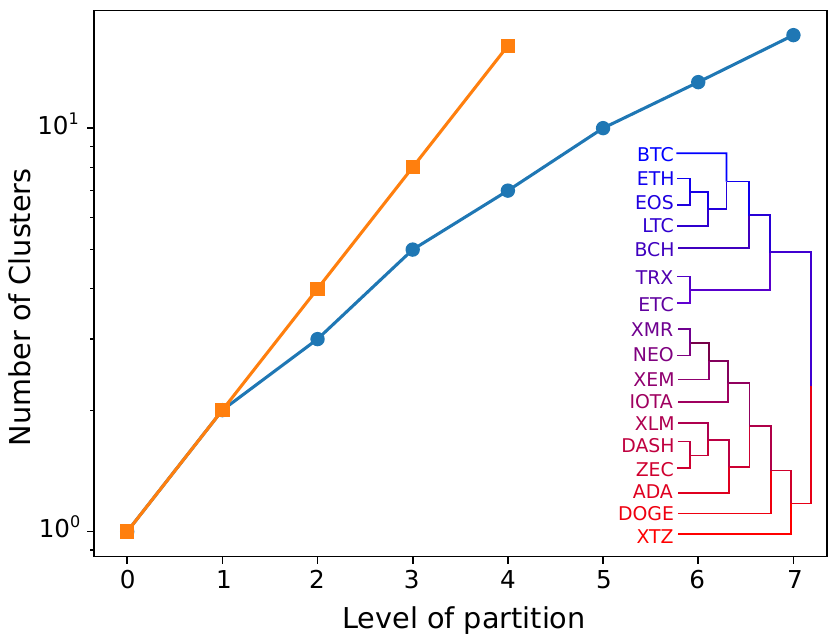}
	\caption{Number of clusters as a function of the level of partition for the
	real data (blue with dots) and for a trivially connected tree (orange with
	squares). The inset shows a tree diagram for the hierarchical structure of
	the assets under study.}
	\label{fig:Cluster}
\end{figure}

The structure of the eigenvectors is typically used to analyze the market. We start with a conjoint set of all assets and then the portfolio is divided according to the sign of the
elements of the eigenvector $\ket{n_{\max}}$ corresponding to the largest eigenvalue $\lambda_{\max}$ \cite{Plerou,Coelho}. 
Since the $\ket{n_{\max}}$ gives the direction of maximum variability in the data, elements with different signs represent positive or negative contributions to this direction.
Each eigenvector after that is associated with variability in a decreasing manner. They can then successively be accessed to divide the market even further until we reach a completely disjoint set. We applied this strategy to the assets under study and obtained the hierarchical tree structure shown in the inset of Fig. \ref{fig:Cluster}.

In greater detail, positive elements from the eigenvector $\mathbf{v}_{\max}$ corresponding to the highest eigenvalue form a partition \#1, whereas negative elements form a partition \#2. Within each of these partitions two new sub-partitions are created according to the sign of the elements of the eigenvector $\mathbf{v}_{\max{}-1}$ corresponding to the second highest eigenvalue. Therefore, partition \#3 is created within partition \#1 with the elements of partition \#1 that are positive in $\mathbf{v}_{\max{}-1}$, whereas partition \#4 is also created within partition \#1 but with the elements of partition \#1 that are negative in $\mathbf{v}_{\max{}-1}$. Similarly, partition number \#5 is created within partition \#2 with elements of partition \#2 that are positive in $\mathbf{v}_{\max{}-1}$. Finally, partition number \#6 is also created within partition \#2 with elements of partition \#2 that are negative in $\mathbf{v}_{\max{}-1}$. The same procedure occurs with elements from the following eigenvectors until there are only single elements.

After performing this partition composition of cryptocurrencies we progressively remove the links in the formed tree structure from
right to left and the number of disjoint clusters are counted. This result
is illustrated in Fig. \ref{fig:Cluster} and is compared with the trivial case (a 3-coordinated Cayley tree)
where a market is uniformly pairwise clustered. The divergence from the
trivial case in our results indicates that the assets under study tend to
strongly bundle together around the same tendency. This is a characteristic of highly correlated systems, which can also be found in spin glasses\cite{sping}.

The difference between the trivial and the measured partition curves results in a function with the form $N\propto e^{p/p_0}$, where $p_0$ is a characteristic partition of the hierarchical tree. This can be used to quantify the correlation level of the system under study with small values implying high correlations. In our case we found $p_0\approx1.0625$.

\subsubsection{Networks}

We further explore the hierarchical structure of the market by calculating
ultrametric ordering, a concept that is borrowed from the theory of frustrated
disordered systems\cite{ultra,Naylor}. In short, an ultrametric space is a metric space where the triangle inequality is restricted to:

\begin{equation}
	d(x,z) \leq \max\{d(x,y),d(y,z)\},
\end{equation}
for any two points $x$ and $y$, and $d(x,z)$ is a distance function between $x$ and $z$. This implies that there is no intermediate points between $x$ and $z$ and a random walk in an ultrametric space is non-ergodic\cite{spingbook}. This is an appropriate way to describe  hierarchical structures since the concept of ultrametricity is tightly related to the concept of hierarchy. Thus, a hierarchical tree can be used to represent an ultrametric set by calculating the distances between nodes. However, computing the distances between all assets creates a complete graph. This
carries redundant information that can be eliminated if one uses a subdominant ultrametric structure.  The latter can be obtained by calculating the the \emph{minimal spanning tree} (MST) \cite{Mant1,Gorski,Schulz}.

A distance between the assets
can be described by the Euclidean distance between their return vectors:

\begin{eqnarray}
		d_{nm}^E  &= ||\mathbf{L}_m-\mathbf{L}_n||\nonumber\\ 
							&= \sqrt{||\mathbf{L}_m||^2+||\mathbf{L}_n||^2-2\mathbf{L}_m\cdot\mathbf{L}_n}\nonumber\\ 
							&= \sqrt{2(1-\Lambda_{nm})}.
\end{eqnarray}
We refer to this metric simply as the Euclidean distance.

Since the structure of the eigenvectors is used to cluster the assets, it is
convenient to also use a \emph{spectral distance}. For this, we create
vectors whose elements are taken from the same entry of different
eigenvectors weighted by the corresponding eigenvalue. The metric is then
the Euclidean distance between these vectors:

\begin{equation}
	d_{nm}^s = \sqrt{\sum_k\lambda_k^2\left[\braket{e_n|k}-\braket{e_m|k}\right]^2}
\end{equation}
We refer to this metric as the spectral distance.

\begin{figure}[h!]
	\centering
	\includegraphics{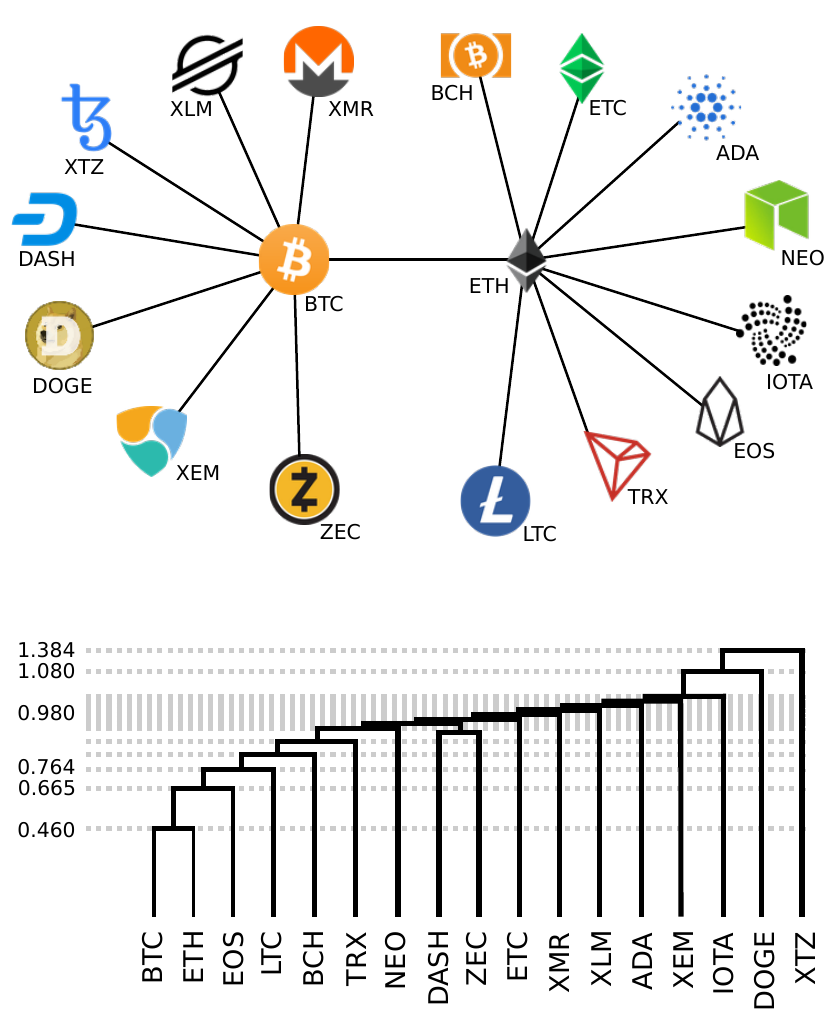}
	\caption{Complex network created using the Euclidean distance (top) and the corresponding dendrogram (bottom).}
	\label{fig:Nets1}
\end{figure}

\begin{figure}[h!]
	\centering
	\includegraphics{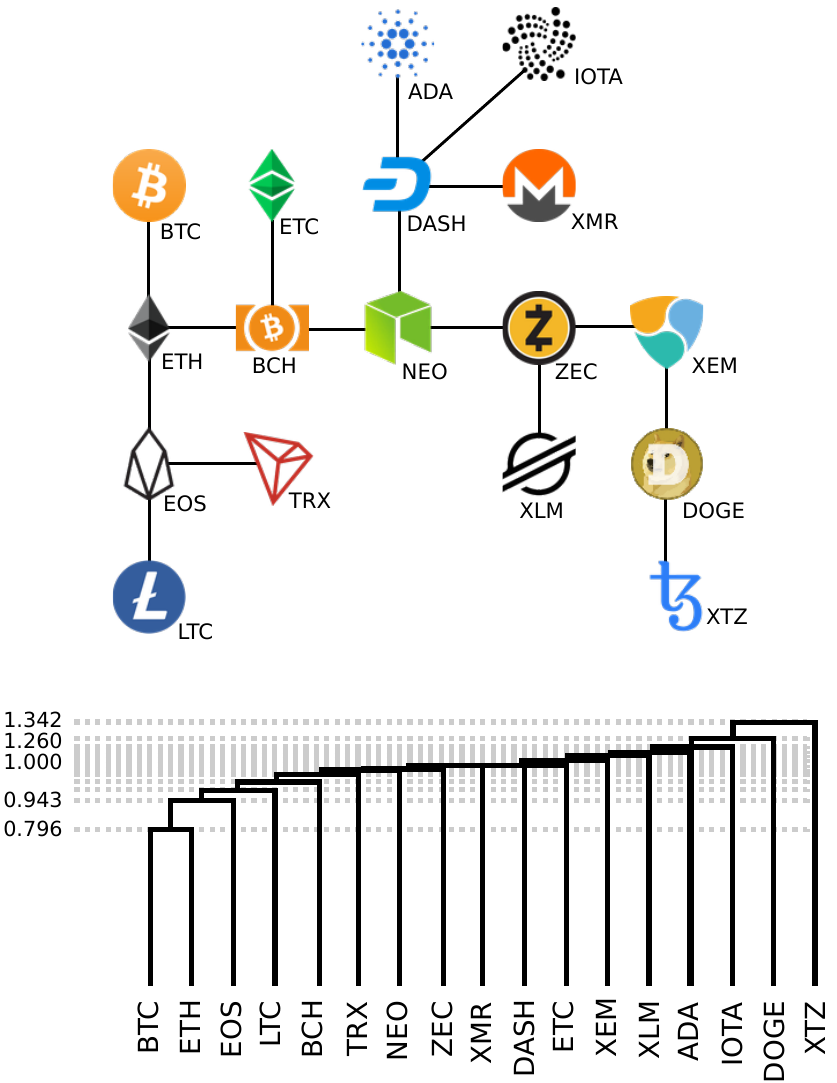}
	\caption{Complex network created using the spectral distance (top) and the corresponding dendrogram (bottom).}
	\label{fig:Nets2}
\end{figure}

We construct such backbone
of the network choosing links such that the total length of the tree is
minimized. We used Prim's algorithm \cite{Prim} to construct the MST shown
in Figs. \ref{fig:Nets1} and \ref{fig:Nets2}. This algorithm consists of connecting the elements
with the smallest distance, then connect the next set of nodes given that no
loop is produced, and repeat this operation until all nodes are connected.

An agglomerative hierarchical clustering algorithm was also used to create hierarchical clusters (dendrograms)\cite{dendro}. In this algorithm, two clusters with the smallest distance are permanently joined together forming a new one. This new cluster $k$ has a distance to the remaining clusters $m$ given by:

\begin{equation}
	d_{km} = \alpha_i d_{im}+\alpha_j d_{jm}+\beta d_{ij}+\gamma|d_{im}-d_{jm}|.
\end{equation}
Here we use a single linkage strategy with $\alpha_i=\alpha_j=\nicefrac{1}{2}$, $\beta=0$, and $\gamma=-\nicefrac{1}{2}$. The agglomeration process continues until a single cluster is present. The resulting dendrograms using both the Euclidean and the spectral distances are shown in Figs. \ref{fig:Nets1} and \ref{fig:Nets2} respectively.

The spanning tree created using the Euclidean distance basically separates the assets in two dissimilar groups, one where BTC is the main hub and another where ETH is the main hub. Its corresponding dendrogram corroborates with this information placing both assets in the bottom of the hierarchy diagram. The dendrogram carries more information. It shows that although EOS, LTC, BCH, TRX, DOGE, and XTZ also form a distinctive hierarchy structure, the remaining assets form a relatively flat correlated structure. 

On the other hand, the spanning tree obtained using the spectral distance shows more details of the correlation structure. For instance, it is clear in the spanning tree that DOGE and XTZ are the most dissimilar assets when compared with BTC and ETH. This happens because the eigenvectors carry information about the relationship between
all elements of the correlation matrix, whereas the Euclidean distance is a
binary distance between assets. Yet, the dendrogram obtained using the spectral distance is very similar to that obtained using the Euclidean distance.

The order of appearance of the assets on both dendrograms have a Pearson correlation coefficient of approximately 70 \% with the market capitalization of the assets. This can be regarded as a stylized fact suggesting that coins with large market capitalization tend to appear on the lowest distances in the dendrograms. Thus, the market seems to select those cryptoassets that are significantly distinctive.

\section{Conclusions}
\label{Section:Conclusions}

We developed a formal parallel between financial assets and quantum systems.
One of the main results is the concept of ``eigenportfolios'' that copy the behavior
of the ground state of a quantum system. We then used this methodology to
study portfolios of 17 digital assets. We computed the correlation
matrix for these assets and compared the density of eigenvalues with those
of a Wishart ensemble and a non-square matrix of random variables drawn from
the Cauchy distribution. Our results indicate that digital assets have a
high explained variance that corresponds to an intermediate behavior
between these two control portfolios. Moreover, the eigenportfolios of digital assets 
have a density of small eigenvalues similar to that of a Wishart ensemble, but a tail
for bigger eigenvalues that is similar to that found for the Cauchy distribution.

We developed a new tool based on the superposition of Gaussian functions to
perform localization analysis on the eigenvectors of the correlation matrix.
Our results indicate that small eigenvalues correspond to highly localized
states as is similarly expected for the Wishart ensemble. Furthermore, a
mild localization behavior similar to that obtained for the Cauchy ensemble
is obtained for larger eigenvalues.

The correlation structure of the eigenportfolios was also probed with ultrametric ordering.
The concept of the minimum spanning tree was used to investigate the
correlation structure of the market of digital assets. We develop a new
metric, based on the spectral characteristics of the correlation matrix to
explore this behavior and show that the corresponding graph is similar to
that obtained by clustering the market according to the elements of its
eigenvectors. Furthermore, this spectral metric produces minimum spanning
trees that carry information similar to that found in the corresponding dendrograms.

The dendrograms may be useful to design portfolios of cryptocurrencies. For instance,
one could create a diverse portfolio choosing assets that dot not strongly correlate
among themselves. Nonetheless, the dendrograms also show that most of the
cryptocurrencies are very correlated, imposing some difficulties in the design of
adequate investment portfolios. It may be possible to lift these degeneracies by
including a virtual perturbation in the correlation matrix (Eq. \ref{eq:eigenp}). 
This can be achieved, for example, by calculating the Markovitz portfolio.

All analyses considered that the correlations between assets are time invariant. 
This however may not be the case in real investment scenarios, and a better
description using the Liouville equation may be required.

\section{Bibliography}
\nocite{*} 
\bibliographystyle{unsrt}
\bibliography{arxiv}

\end{document}